\documentclass[useAMS,usenatbib]{mn2e}

\usepackage{amsmath,array,amsfonts}
\usepackage{amssymb}
\usepackage{graphicx}
\usepackage{natbib}
\usepackage{pstricks}

\newcommand{\ltsima} {$\; \buildrel < \over \sim \;$}
\newcommand{\gtsima} {$\; \buildrel > \over \sim \;$}
\newcommand{\lta} {\lower.5ex\hbox{\ltsima}}
\newcommand{\gta} {\lower.5ex\hbox{\gtsima}}

\title[WMAP constraints on $f_{NL}$]{Wilkinson Microwave Anisotropy Probe 7-yr constraints on $f_{NL}$ with a fast wavelet estimator}

\author[B. Casaponsa et al.]{B. Casaponsa,$^1$ $^2$\thanks{e-mail:
casaponsa@ifca.unican.es} R. B. Barreiro,$^1$ A. Curto,$^1$ E. Mart\'{\i}nez-Gonz\'alez,$^1$ P. Vielva$^1$ \\ 
$^1$     Instituto de F\'isica de Cantabria, CSIC-Universidad de Cantabria, Avda. de los Castros s/n, 39005 Santander, Spain.\\
$^2$     Dpto. de F\'isica Moderna, Universidad de Cantabria, Avda. los Castros s/n, 39005 Santander, Spain.}
\date{Accepted  Received ; in original form }

\begin{document}

\maketitle

\begin{abstract}
A new method to constrain the local non-linear coupling parameter $f_{NL}$ based on a fast wavelet decomposition is presented. Using a multiresolution wavelet adapted to the HEALPix pixelization, we have developed a method that is $\sim10^{2}$ times faster than previous estimators based on isotropic wavelets and $\sim10^{3}$ faster than the KSW bispectrum estimator, at the resolution of the Wilkinson Microwave Anisotropy Probe (WMAP) data. The method has been applied to the WMAP 7-yr V+W combined map, imposing constraints on $f_{NL}$ of $-69<f_{NL}<65$ at the 95 per cent CL. This result has been obtained after correcting for the contribution of the residual point sources which has been estimated to be $\Delta f_{NL}=7\pm6$. In addition, a Gaussianity analysis of the data has been carried out using the third order moments of the wavelet coefficients, finding consistency with Gaussianity. Although the constrainsts imposed on $f_{NL}$ are less stringent than those found with optimal estimators, we believe that a very fast method, as the one proposed in this work, can be very useful, especially bearing in mind the large amount of data that will be provided by future experiments, such as the Planck satellite. Moreover, the localisation of wavelets allows one to carry out analyses on different regions of the sky. As an application, we have separately analysed the two hemispheres defined by the dipolar modulation proposed by \citet{Hoftuft2009}. We do not find any significant asymmetry regarding the estimated value of $f_{NL}$ in those hemispheres.


\end{abstract}
\section{Introduction}
The cosmic microwave background (CMB) is one of the pillars that provide support to the Big Bang theory.
The fluctuations of the CMB naturally arise in an inflationary scenario. The understanding of this very early stage of the history of the Universe is a challenging issue for the scientific community due to the implications on large scale structure formation and fundamental particle physics at high energies. A large number of inflationary models have been proposed in the literature (for an overview see for instance \citealt{Lyth2008}) but the task of testing such scenarios is not trivial, and there is the need of new experiments and powerful statistical tools to discriminate among them. In this sense, the statistical properties of the CMB temperature anisotropies are a source of information about the processes that have generated the primordial fluctuations. In particular, the standard, slow roll, single field inflationary model predicts a nearly Gaussian distribution of the CMB temperature anisotropies, while alternative models may introduce a certain level of non-Gaussianity in the CMB. A convenient parametrization valid for a large set of non-standard inflationary models which includes the quadratic corrections of the primordial curvature perturbation is \citep{Salopek1990,Gangui1994,Verde2000,Komatsu2001}:
\begin{equation}
\phi(r)=\phi_{L}(r)+f_{NL} \left [\phi_{L}^{2}-<\phi_{L}^{2}>\right ],
\label{eq:fnl}
\end{equation}where $\phi_{L}$ are Gaussian linear perturbations and $f_{NL}$ characterises the amplitude of the non-linear contribution in real space. This local form appears in non-standard multi-field inflationary models \citep{Babich2004,Komatsu2009}. For a complete review on non-Gaussianity due to inflationary models see \citet{Bartolo2004}. In addition to inflationary models, there are other alternative scenarios that can be constrained, such as the ekpyrotic model where a negative value of $f_{NL}$ is expected \citep{Lehners2010}. Moreover, there are other processes that can introduce deviations from Gaussianity in the third order moments (as foreground contamination, non-linear gravitational effects, topological defects, etc).

Since the quadratic parametrization was proposed, an important effort has been made to set observational constraints on local $f_{NL}$ with a wide variety of methods including the bispectrum \citep{Yadav2008,Smith2009,Komatsu2010}, wavelet-based methods \citep{Cayon2003,Mukherjee2004, Curto2009a,Curto2009b, Pietrobon2009, Rudjord2009}, Minkowski functionals \citep{Hikage2008} or the N-pdf \citep{Vielva2009,Vielva2010}. Most of these works find that the data are compatible with $f_{NL}=0$, but the constraints are not yet sufficiently tight to discriminate among a large set of inflationary models. The best current limit is given by \citet{Komatsu2010} and is $-10 < f_{NL} < 74$ at the 95 per cent confidence level. These constraints have been obtained with a bispectrum estimator, which is computationally very demanding. However, \citet{Curto2010} have recently shown that an estimator based on the SMHW can provide constraints on $f_{NL}$ as stringent as the optimal estimator, the bispectrum, while reducing considerably the CPU time. With the arrival of new data from high resolution experiments such as the ESA Planck satellite\footnote{http://www.rssd.esa.int/index.php?project=planck} \citep{Tauber2010}, it becomes even more important the availability of even faster and simpler methods. With this aim we present the application for CMB of a wavelet adapted to the HEALPix pixelization similar to the tool proposed by \citet{Shahram2007}.\\
\\
The paper is organised as follows. In Section~\ref{sec:wavelet} we introduce the HEALPix wavelet decomposition. In Section~\ref{sec:methodology} the method to constrain the $f_{NL}$ parameter as well as the proposed Gaussianity test are described. In Section~\ref{sec:results} we present the results of the application of this technique to the WMAP 7-yr data. Finally, our conclusions are summarised in Section~\ref{sec:conclusions}.

%
\section{The HEALPix wavelet}
\label{sec:wavelet}

A large set of different wavelets have been used in the astrophysics literature. In particular, different spherical wavelets have been applied to CMB Gaussianity analysis during the last decade, including the spherical Haar wavelet (SHW, \citealt{Barreiro2000}), the SMHW (\citealt{Martinez2002,Vielva2004,Mukherjee2004,Cruz2005,Curto2009b}), elliptical SMWH \citet{McEwen2005}, directional spherical wavelets \citep{McEwen2005,McEwen2006,McEwen2008} and needlets \citep{Pietrobon2009,Rudjord2009,Cabella2010}. For a review on wavelet applications to cosmology see \citealt{McEwen2007}.

\begin{figure*}
 \center 
\psset{unit=1.3cm}
\begin{pspicture}(0,0)(12,6)
\psgrid[subgriddiv=2,gridlabels=0](0,2)(2,4)

\psgrid[subgriddiv=1,gridlabels=0](4,1)(6,3)
\psline[linecolor=black](4,3.5)(4,5.5)(6,5.5)(6,3.5)(4,3.5) 
\psline[linecolor=black](5,3.5)(5,5.5)
\psline[linecolor=black](4,4.5)(6,4.5) 
\psline[linecolor=black,linewidth=0.01pt,linecolor=gray](4.5,3.5)(4.5,5.5)
\psline[linecolor=black,linewidth=0.01pt,linecolor=gray](5.5,3.5)(5.5,5.5) 
\psline[linecolor=black,linewidth=0.01pt,linecolor=gray](4,4)(6,4)
\psline[linecolor=black,linewidth=0.01pt,linecolor=gray](4,5)(6,5) 
\psline[linecolor=black](8,2.5)(8,4.5)(10,4.5)(10,2.5)(8,2.5) 
\psline[linecolor=black](9,2.5)(9,4.5)
\psline[linecolor=black](8,3.5)(10,3.5) 
 \psline[linecolor=black](8,0)(8,2)(10,2)(10,0)(8,0) 
\uput[0](0,3.75){$x_{00}$}
\uput[0](0.5,3.75){$x_{01}$}
\uput[0](0,3.25){$x_{02}$}
\uput[0](0.5,3.25){$x_{03}$}
\uput[0](1,3.75){$x_{10}$}
\uput[0](1.5,3.75){$x_{11}$}
\uput[0](1,3.25){$x_{12}$}
\uput[0](1.5,3.25){$x_{13}$}
\uput[0](0,2.75){$x_{20}$}
\uput[0](0.5,2.75){$x_{21}$}
\uput[0](0,2.25){$x_{22}$}
\uput[0](0.5,2.25){$x_{23}$}
\uput[0](1,2.75){$x_{30}$}
\uput[0](1.5,2.75){$x_{31}$}
\uput[0](1,2.25){$x_{32}$}
\uput[0](1.5,2.25){$x_{33}$}
\uput[0](4,5.25){$d_{00}$}
\uput[0](4.5,5.25){$d_{01}$}
\uput[0](4,4.75){$d_{02}$}
\uput[0](4.5,4.75){$d_{03}$}
\uput[0](5,5.25){$d_{10}$}
\uput[0](5.5,5.25){$d_{11}$}
\uput[0](5,4.75){$d_{12}$}
\uput[0](5.5,4.75){$d_{13}$}
\uput[0](4,4.25){$d_{20}$}
\uput[0](4.5,4.25){$d_{21}$}
\uput[0](4,3.75){$d_{22}$}
\uput[0](4.5,3.75){$d_{23}$}
\uput[0](5,4.25){$d_{30}$}
\uput[0](5.5,4.25){$d_{31}$}
\uput[0](5,3.75){$d_{32}$}
\uput[0](5.5,3.75){$d_{33}$}
\uput[0](4.20,2.5){\Large$\bar{x}_{00}$}
\uput[0](5.20,2.5){\Large$\bar{x}_{01}$}
\uput[0](4.2,1.5){\Large$\bar{x}_{02}$}
\uput[0](5.2,1.5){\Large$\bar{x}_{03}$}
\uput[0](8.2,4){\Large$d_{00}$}
\uput[0](9.2,4){\Large$d_{01}$}
\uput[0](8.2,3){\Large$d_{02}$}
\uput[0](9.2,3){\Large$d_{03}$}
\uput[0](8.55,1){\Huge$\bar{x}\scriptscriptstyle_{00}$}

\psline{->}(2.5,3.25)(3.5,4)
\rput[bl]{38}(2.5,3.35){$d_{ij}=x_{ij}-\bar{x}_{0i}$}
\psline{->}(2.5,3)(3.5,2.25)
\rput[bl]{-38}(2.15,2.40){$\bar{x}_{0i}=\displaystyle\sum_{j=0}^{3}\frac{x_{ij}}{4}$}
\psline{->}(6.5,2.25)(7.5,3)
\rput[bl]{38}(6.5,2.35){$d_{0i}=x_{0i}-\bar{x}_{00}$}
\psline{->}(6.5,2)(7.5,1.25)
\rput[bl]{-38}(6.15,1.40){$\bar{x}_{00}=\displaystyle\sum_{i=0}^{3}\frac{\bar{x}_{0i}}{4}$}
\psline{->}(10.25,1.10)(10.75,1.65)
\psdot[dotsize=2pt](11,1.65)
\psdot[dotsize=2pt](11.25,1.65)
\psdot[dotsize=2pt](11.5,1.65)
\psline{->}(10.25,0.9)(10.75,0.4)
\psdot[dotsize=2pt](11,0.4)
\psdot[dotsize=2pt](11.25,0.4)
\psdot[dotsize=2pt](11.5,0.4)
\end{pspicture}
\caption{Diagram of the construction of the approximation and detail coefficients. Approximation coefficients are computed as the average of the four daughter pixels. Detail coefficients are computed as the subtraction of that average from the original pixels and are represented by \textit{d}.}
 \label{fig:wavelet_diagram}
 \end{figure*}
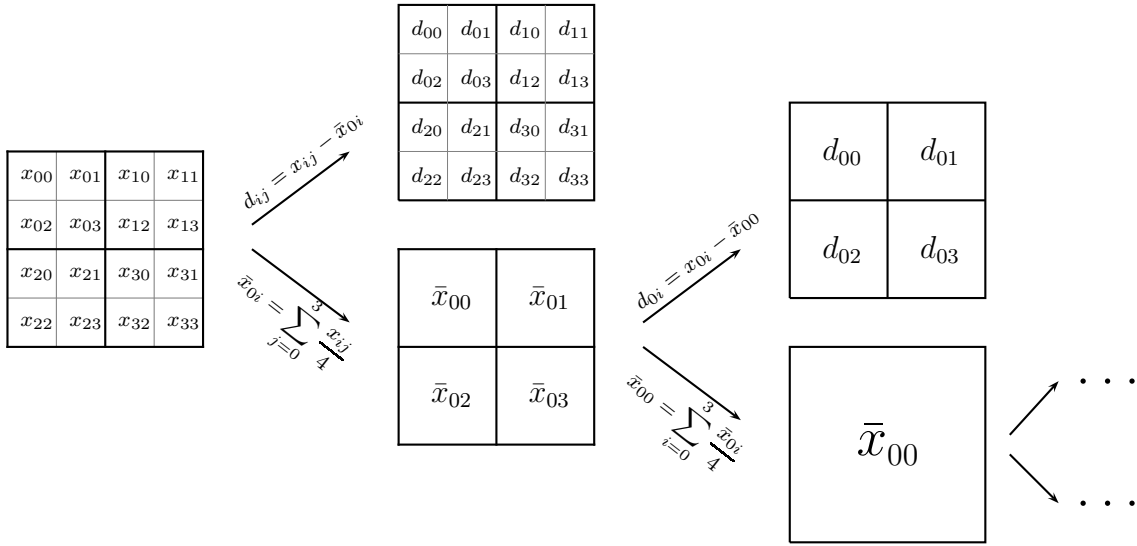

In this work, we will present an application using the so-called HEALPix wavelet (HW). In the previous work of \citep{Shahram2007} a linear operator is applied to the HW to obtain wavelet coefficients corresponding to vertical, horizontal and diagonal orientations. This operation leads to a wavelet coefficients without redundancy, obtaining a number of wavelet coefficients (details plus approximation) equal to the number of original pixels. However, we have kept the HW with its intrinsic redundancy for three main reasons: first, to improve the computational time, second to obtain a wavelet decomposition as isotropic as the HW allows \footnote{ It is worth mentioning that HW detail coefficients help to highlight the isotropy properties of the field as compared to the directional oriented details of the SHW. This is important because the local non-Gaussianities are expected to be isotropic.}, and third, because, as it is shown later, redundancy helps to improve the sensitivity in the detection of $f_{NL}$.  Similarly to the SHW, the HW is a discrete, orthogonal wavelet, adapted to a hierarchical pixelization (such as HEALPix\footnote{http://healpix.jpl.nasa.gov/}, \citealt{Gorski2005}),  whereas the SMHW is a continuous, non-orthogonal wavelet and does not have a hierarchical decomposition structure. The HW presents an optimal space localization, while the scale localization is not as good as that of the SMHW. It is important to point out that, in the case of the SMHW and of needlets, a transformation of the data into spherical harmonic space is required. However, this is not the case for the HW and, therefore, the computational cost is significantly reduced.  \\
The resolution of a HEALPix map is characterised by the $N_\mathrm{side}$ parameter, such that the number of pixels in which the sphere is divided corresponds to $N=12 N_\mathrm{side}^2$. $N_\mathrm{side}$ can only take powers of two as values. The HW decomposes the temperature map at resolution $J$, where $N_\mathrm{side}=2^{J}$, in wavelet coefficient maps at all the allowed HEALPix resolutions down to the lower considered resolution $j_{0}$. The wavelet functions are defined as follows:
\begin{gather}
\Psi_{0,j,k}(x)=\varphi _{j+1,k_{0}}(x)-\frac{\varphi_{j,k}(x)}{4}\\ \nonumber
\Psi_{1.j,k}(x)=\varphi _{j+1,k_{1}}(x)-\frac{\varphi_{j,k}(x)}{4}\\  \nonumber
\Psi_{2,j,k}(x)=\varphi _{j+1,k_{2}}(x)-\frac{\varphi_{j,k}(x)}{4}\\ \nonumber
\Psi_{3,j,k}(x)=\varphi _{j+1,k_{3}}(x)-\frac{\varphi_{j,k}(x)}{4}
\end{gather}
where $\varphi(x)_{j,k}$ is the scaling function 
\begin{equation}
\varphi(x)_{j,k}=\left\{ \begin{array}{ll}
                1 & if\; x\in P_{j,k}\\
                0 & \mathrm{otherwise}\;, \\
                \end{array} \right.  
\end{equation}
and $P_{j,k}$ is the pixel at position $k$ at resolution $j$, which at the next higher resolution is divided into four daughter pixels $P_{j+1,k_{0}}$, $P_{j+1,k_{1}}$, $P_{j+1,k_{2}}$, $P_{j+1,k_{3}}$.\\
\\
The wavelet decomposition of a temperature map can be written in terms of the basis functions and a set of coefficients:
\begin{gather}
 \frac{\Delta T}{T}(x_{i})=\sum_{k=0}^{N_{j_0}-1}\lambda_{j_{0},k}\varphi_{j_{0},k}(x_{i})+\nonumber\\ 
+\sum_{j=j_{0}}^{J}\sum_{m=0}^{3}\sum_{k=0}^{N_{j}-1}\gamma_{m,j,k}\Psi_{m,j,k}(x_{i})\;, \label{eq:reconstruccion}
\end{gather}
where $N_{j}$ is the number of pixels at resolution $j$. $\lambda_{j,k}$ and $\gamma_{m,j,k}$ are the approximation and detail coefficients respectively. From a practical point of view, to perform the decomposition, we start with the original resolution, i.e. $j=J$. At this resolution, the approximation coefficients $\lambda_{J,k}$ correspond to the pixels of the original tempature map. The approximation coefficients at the next resolution are simply obtained by degrading the map to the inmediatly lower resolution (i.e., by averaging the corresponding four daughter pixels):
\begin{equation}
\lambda_{j,k}=\frac{1}{4}\sum_{i=0}^{3}\lambda_{j+1,k_{i}} \;,
\end{equation}
On the other hand, the detail coefficients at resolution $j+1$ are simply obtained by subtracting the approximation at resolution $j$ from the approximation at resolution $j+1$. Thus, the detail coefficients are defined as:
\begin{gather}
\gamma_{0,j,k}=\lambda_{j+1,k_{0}}-4\lambda_{j,k} \\ \nonumber
\gamma_{1,j,k}=\lambda_{j+1,k_{1}}-4\lambda_{j,k}\\ \nonumber
\gamma_{2,j,k}=\lambda_{j+1,k_{2}}-4\lambda_{j,k}\\ \nonumber
\gamma_{3,j,k}=\lambda_{j+1,k_{3}}-4\lambda_{j,k} 
\end{gather}
A schematic diagram of how to obtain the approximation and detail coefficients is given in Fig.~\ref{fig:wavelet_diagram}.

\section{Methodology}
\label{sec:methodology}
The main purpose of this work is to constrain the parameter $f_{NL}$ defined in Eq.~(\ref{eq:fnl}) using the WMAP-7yr data\footnote{The data are available at the LAMBDA web page: http://lambda.gsfc.nasa.gov/}. For this analysis, we only consider the (foreground reduced) V and W channels, since they are less afected by foreground contamination. A single CMB map is obtained through a noise-weighted linear combination of the V and W receivers. The KQ75 mask (which covers around a 29 per cent of the sky) is applied subsequently.

In order to calibrate our estimator, we need both Gaussian and non-Gaussian simulations. To generate the Gaussian simulations, we compute the power spectrum that best fits the WMAP-7yr data accordingly to the parameters estimated by \citet{Komatsu2010}. For this purpose we use the On-line tool CAMB \citep{Lewis2000}. We then apply the corresponding beam and pixel functions to simulate the data at each of the considered receivers (2 for V and 4 for W). A Gaussian noise realisation is subsequently added to the CMB maps with a variance per pixel given by $\frac{\sigma_{0}}{N_{obs}}$, where $\sigma_{0}$ is the detector sensitivity of each of the receivers and $N_{obs}$ is the number of observations at each pixel. Finally the six maps are combined in the same way as the data.

Regarding the non-Gaussian simulations, we have used the 1000 simulations generated by \citet{Elsner2009} that are publicly available\footnote{http://planck.mpa-garching.mpg.de/cmb/fnl-simulations/}. The previous authors provide the harmonic coefficients for the Gaussian and non-Gaussian parts of the simulation. A non-Gaussian simulation with a given value of $f_{NL}$ is then constructed as:
\begin{equation}
a_{lm}=a_{lm}^{(G)}+f_{NL}a_{lm}^{(NG)}\;, \label{eq:alm_fnl} 
\end{equation} 
where we have normalized $a_{lm}^{(G)}$ and $a_{lm}^{(NG)}$\footnote{The amplitude of the $a_{lm}^{NG}$ has been corrected by a factor of $\frac{3}{5}$ as indicated by the authors} to the power spectrum that best fits the WMAP-7yr data, and that was used for the Gaussian simulations (the original simulations were obtained using the WMAP 5-yr power spectrum). Again, we construct the maps for the V and W receivers, applying the corresponding beam and pixel transfer functions and adding the appropriate level of noise. Finally a single V+W combined map is constructed for each non-Gaussian simulation.

\subsection{Cubic statistics}
\label{subsec:statistics}
In this section we define the third order moments of the wavelet coefficients that are used to constrain $f_{NL}$. Similar statistics have been used in other previous works \citep{Curto2009b,Rudjord2009}.\\
We perform the wavelet decomposition of the considered map starting at resolution $N_\mathrm{side}=512$ $(J=9)$ and down to $N_\mathrm{side}=2$ $(j_{0}=1)$ (when using a higher value of $j_{0}$ we are losing efficiency whereas for $j_{0}=0$ the results are not significantly improved while the computational time increases by a 30\%). We obtain 8 detail maps and 1 approximation map. In addition, we also include in the analysis the original map and the 8 intermediate approximation maps (which are obtained during the wavelet decomposition to construct the detail coefficients). Although, in principle, these additional approximation maps contain redundant information, they seem to provide additional information regarding the third order statistics, since a larger number of third order combinations can be constructed. In fact, we have tested that with the inclusion of the approximation maps and the original map, the results are improved by a 30\%.  Therefore, we have a total of 18 maps for each analysed signal. 
The statistics are constructed as the third order moments of all the possible combinations of these 18 maps, where the coefficients are weighted to take into account the presence of a mask. In order to calculate these weights, one performs the wavelet decomposition of the considered mask (that has zeros in the masked pixels and ones in the rest). The wavelet coefficients of the mask at each detail and approximation scale are used to construct the weight $w_{j}(i)$ of the coefficient at position $i$ at resolution $j$. This makes sense if one bears in mind how the wavelet decomposition is performed. For instance, to construct the approximation map at resolution $J-1$ at a given position $i$, the four daughter pixels at resolution $J$ have to be averaged. If the four pixels are unmasked, this corresponds to a weight of 1 in the original map and also in the approximation map at position $i$. However, if one of the orignal pixels is masked, this pixel would have a weight of zero, and the average would be done only over three pixels. Thus the weight of the corresponding approximation coefficient would be 3/4. Therefore, this weighting scheme takes into account the fact that different coefficients contain different amount of information, depending on the considered mask. Also, contrary to the case of other wavelet estimators, this means that the mask does not need to be extended but, in fact, it is reduced when increasing the scale. This is due to the fact that a larger pixel is kept for the analysis, with the appropriate weight, if at least one of the daughter pixels was unmasked.

The third order statistics are then defined as:
\begin{equation}
 S_{jkl}=\frac{1}{\displaystyle\sum_{i=0}^{N_{l}-1}W_{jkl}(i)}\sum_{i=0}^{N_{l}-1}\frac{W_{jkl}(i)\epsilon_{i,j}\epsilon_{i,k}\epsilon_{i,l}}{\sigma_{j}\sigma_{k}\sigma_{l}},  \label{eq:weight_stat}
\end{equation}
where $\epsilon_{i,j}=y_{i,j}-\mu_{j}$ and $y_{i,j}$ are the wavelet coefficients maps at position $i$ at resolution $j$. Note that $j$ goes from $j_{0}$ to $J$, $k$ goes from $j$ to $J$ and $l$ goes from $k$ to $J$. $\mu_{j}$ and $\sigma_{j}$ are the weighted mean and the dispersion for the map at resolution $j$. $W_{jkl}(i)$ is the weight associated to the wavelet coefficients at position $i$ and scales $j,k,l$ and is given by:
\begin{equation}
W_{jkl}(i)=\sqrt[3]{w_{j}(i)w_{k}(i)w_{l}(i)}\;.
\end{equation}

Note that some of these statistics are redundant (linearly dependent between them), so we restrict our analysis to the set of non-rendudant statistics, which gives a total of $n_\mathrm{stat}=$232 quantities.

The process for computing these statistics requires $\sim N\times n_\mathrm{stat}$ number of operations, where N is the number of pixels and $n_\mathrm{stat}$ the number of statistics computed. This number is significantly lower than that of the full bispectrum that needs $N^{\frac{5}{2}}$ operations. Using the KSW algorithm presented in \citet{Komatsu2005} the number of operations is reduced to $\sim rN^{\frac{3}{2}}$, where r is the number of sampling points (of the order of 100). On the other hand, the SMHW scales as $\sim n_{s}N^{\frac{3}{2}}$, where $n_{s}$ is the number of scales involved (of the order of 10). Thus, at WMAP resolution ($N_\mathrm{side}=512$ and $N\sim3\times10^{6}$) we have that the method presented in this work is $10^{2}$ times faster than the SMHW, $10^{3}$ times faster than KSW bispectrum estimator and $10^{7}$ faster than the general bispectrum estimator.

%
\subsection{Gaussianity test and $f_{NL}$ constraints}
We first perform a Gaussianity test in order to probe whether the data is compatible with Gaussianity using the $\chi^{2}$ estimator:
\begin{equation}
\chi^{2}=\sum_{i,j=1}^{n_\mathrm{stat}}(v_{i}-\langle v_{i}\rangle)C^{-1}_{ij}(v_{j}-\langle v_{j}\rangle)\;, \label{eq:chi_modelo0}
\end{equation}
where $v_{i}$ is the vector of the third order statistics computed from the considered map (to simplify notation, hereinafter we define  $v_{1}\equiv S_{111}, v_2 \equiv S_{112}, ...$). $\langle v_{i}\rangle$ and $C_{ij}$ are the mean and covariance matrix of the statistics obtained with 10000 Gaussian simulations.
To perform the Gaussianity test, the value of the $\chi^2$ is computed for the WMAP data, and compared to the distribution of the estimator obtained from an additional set of 1000 Gaussian simulations.\\

%
The second analysis that has been performed is the estimation of $f_{NL}$ from the data. As the wavelet decomposition is linear, we can obtain the wavelet coefficients from the Gaussian and non-Gaussian parts separately. Thus, the wavelet coefficients for a given value of $f_{NL}$ are given by
\begin{equation}
y_{i}=y_{i}^{(G)}+f_{NL}y_{i}^{(NG)}\;.
\end{equation}
Taking into account that $y^{(NG)}$ are around 4 orders of magnitude smaller than $y^{(G)}$, when we compute $<y^{3}>$ the NG high-order terms can be neglected and it can be shown that $f_{NL}$ is proportional to the wavelet estimators, as it is also the case for other statistics (such as the bispectrum):
\begin{equation}
v_{i}=a_{i} f_{NL}\;, \label{eq:model}
\end{equation}
where $a_{i}$ can be computed from simulations with a simple linear regression. 

In order to estimate the $f_{NL}$ parameter, we perform a $\chi^{2}$ minimisation. In particular, $\chi^{2}(f_{NL})$ is defined as follows 
\begin{equation}
\chi^{2}=\sum_{i,j=1}^{n_\mathrm{stat}}(v_{i}-\langle v_{i}\rangle _{f_{NL}})C^{-1}_{ij}(f_{NL})(v_{j}-\langle v_{j}\rangle _{f_{NL}})\;, \label{eq:chi_modelo1}
\end{equation} 
where $\langle v_{i}\rangle_{f_{NL}}$ is the mean of the statistics for a given value of $f_{NL}$ obtained from the 1000 non-Gaussian simulations and $C_{ij}(f_{NL})$ is the corresponding covariance matrix. For $f_{NL}<< 1000$ is reasonable to use the approximation $C_{ij}(f_{NL})\simeq C_{ij}$, where $C_{ij}$ is the covariance matrix for the Gaussian case. 

Error bars on the parameter estimation at different confidence levels are found using the Gaussian simulations. We also compute the minimum variance in a semi-analytical manner. It is well known that the diagonal of the inverse of the Fisher matrix provides an estimation of the variance of the parameters. In order to estimate the Fisher matrix, we approximate the distribution of the statistics by a Gaussian. Using this approximation and taking into account Eqs.~(\ref{eq:model}) and (\ref{eq:chi_modelo1}), the variance from the Fisher Matrix can be written as:
\begin{equation}
\displaystyle
\sigma^{2}=\frac{1}{\displaystyle\sum_{i,j}^{n_\mathrm{stat}}a_{i}C^{-1}_{ij}a_{j}}\;. \label{eq:sigma_fisher}
\end{equation}
In practice, the distribution of the statistics do not follow a perfect Gaussian distribution. Therefore, this variance can be seen as a lower limit to the true underlying variance.

%
\section{Results}
\label{sec:results}
In this section, we present the analysis of the WMAP-7yr V+W combined map. On the one hand, we analyse the compatibility of the data with Gaussianity using the cubic statistics defined in Eq.~(\ref{eq:weight_stat}) and the estimator presented in Eq.~(\ref{eq:chi_modelo0}). On the other hand, we compute the best-fit $f_{NL}$ parameter from the data by minimizing Eq.~(\ref{eq:chi_modelo1}). Error bars are set using simulations. In addition, we also present a study of the contribution of the point sources to the estimated $f_{NL}$ value and of the variation of $f_{NL}$ estimated from two independent hemispheres (defined by \citealt{Hoftuft2009}).

\subsection{Gaussianity test}
\begin{figure}
 \center 
\begin{tabular}{c c} 
\includegraphics[width=9cm,height=6cm, angle = 0] {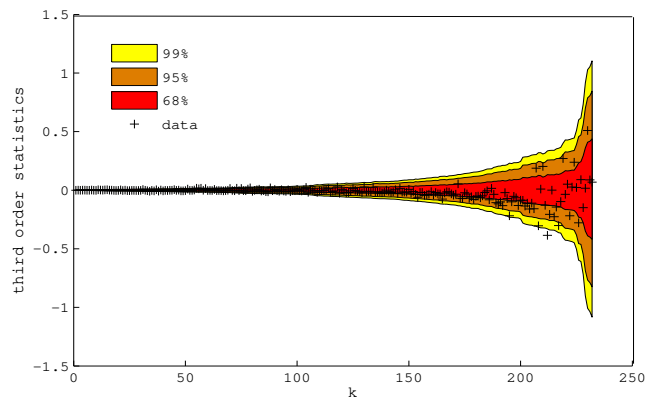} 
\end{tabular} 
\caption{The cubic statistics $v_{i}$ from WMAP-7y V+W data are shown. Shadow areas correspond to the 68, 95 and 99 per cent confidence levels of the distribution obtained from 1000 Gaussian simulations. The statistics have been plotted from lower to higher variance.} 
 \label{fig:test_gaussianity}
 \end{figure}

As explained in Section~\ref{subsec:statistics}, we have considered a total of 232 cubic statistics, constructed from 18 maps at 9 different scales. Fig.~\ref{fig:test_gaussianity} shows the value of $v_i$ for the WMAP 7-yr V+W data, after applying the KQ75 mask. 
The plot does not show any obvious deviation from Gaussianity. To further study the consistency of the data with Gaussianity, we also perform the $\chi^{2}$ test defined in Eq.~(\ref{eq:chi_modelo0}). From 1000 Gaussian simulations, we estimate the distribution of this quantity, finding a mean value of $\left\langle\chi^{2}\right\rangle=233$, very close to the number of degrees of freedom (232). The value of the dispersion is 69, larger than expected for a $\chi^2$ distribution with the considered degrees of freedom. However, this may be explained by the fact that the distribution of the different statistics are not purely Gaussian. For the WMAP data, we find $\chi^{2}_{data}=434$  with a cumulative probability of $P(\chi^{2}\leq\chi^{2}_{data})=0.96$. Although the result indicates that the WMAP data is some how in the tail of the distribution, the $\chi^2$ value is not large enough to claim a deviation from Gaussianity.

\subsection{Constraints on the $f_{NL}$ parameter}

\begin{figure}
 \center 
\includegraphics[width=9cm,height=6cm, angle = 0] {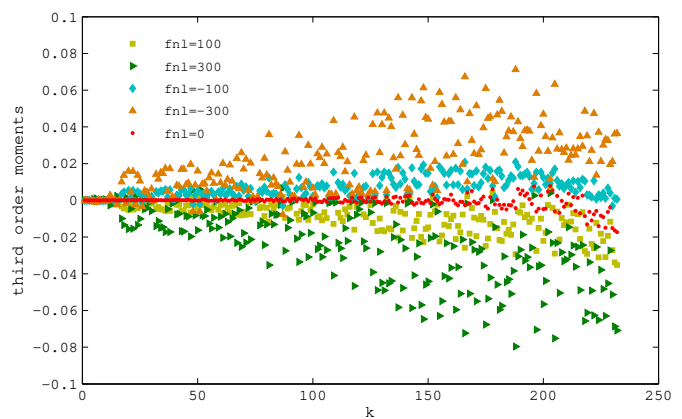} 
\caption{Mean values of the cubic statistics $v_{i}$ from 1000 non-Gaussian simulations with $f_{NL}=0,\pm100,\pm300$.} 
 \label{fig:3rd_ord_moments}
 \end{figure}

We have also performed an estimation of the non-linear parameter $f_{NL}$. As already mentioned, for this analysis we have used the 1000 non-Gaussian simulations provided by \cite{Elsner2009}. In Fig.~\ref{fig:3rd_ord_moments}, the mean of the cubic statistics derived from simulations with $f_{NL}=0,\pm100,\pm300$ is presented. 
It can be seen that, as stated in Eq.~(\ref{eq:model}), the statistics are proportional to the value of $f_{NL}$. 

After obtaining the cubic statistics for the WMAP-7yr data and minimising the $\chi^2$ given by Eq.~(\ref{eq:chi_modelo1}), we estimate that the best-fit value of $f_{NL}$ is 6. Using Gaussian simulations, we find that the contraints for the parameter are $-28 < f_{NL} < 40$  at the 68 per cent confidence level and $-62 < f_{NL} < 72$ at the 95 per cent confidence level. It is also interesting to point out the agreement between the dispersion computed semi-analytically through the Fisher matrix (Eq.~\ref{eq:sigma_fisher}) and that obtained from Gaussian simulations, which are both estimated to be around 34.

Although the constraints provided by the HW are less stringent than those found with optimal estimators (such as the bispectrum or the SMHW), they are similar or even better than those obtained by other mehtods such as needlets \citep{Pietrobon2009,Rudjord2009}, the Minkowski functionals \citep{Hikage2008} or the N-pdf \citep{Vielva2009}. Moreover, as already pointed out, our estimator is significantly faster than all the previously mentioned methods, providing a very valuable tool, especially for future high resolution experiments such as Planck. It is also interesting to point out that we find a more symmetric constraint around zero than those obtained, for instance, by \citet{Komatsu2010} or \citet{Curto2010}.

\begin{figure*}
 \centering
\begin{tabular}{c c}
\includegraphics[width=9cm,height=6cm]{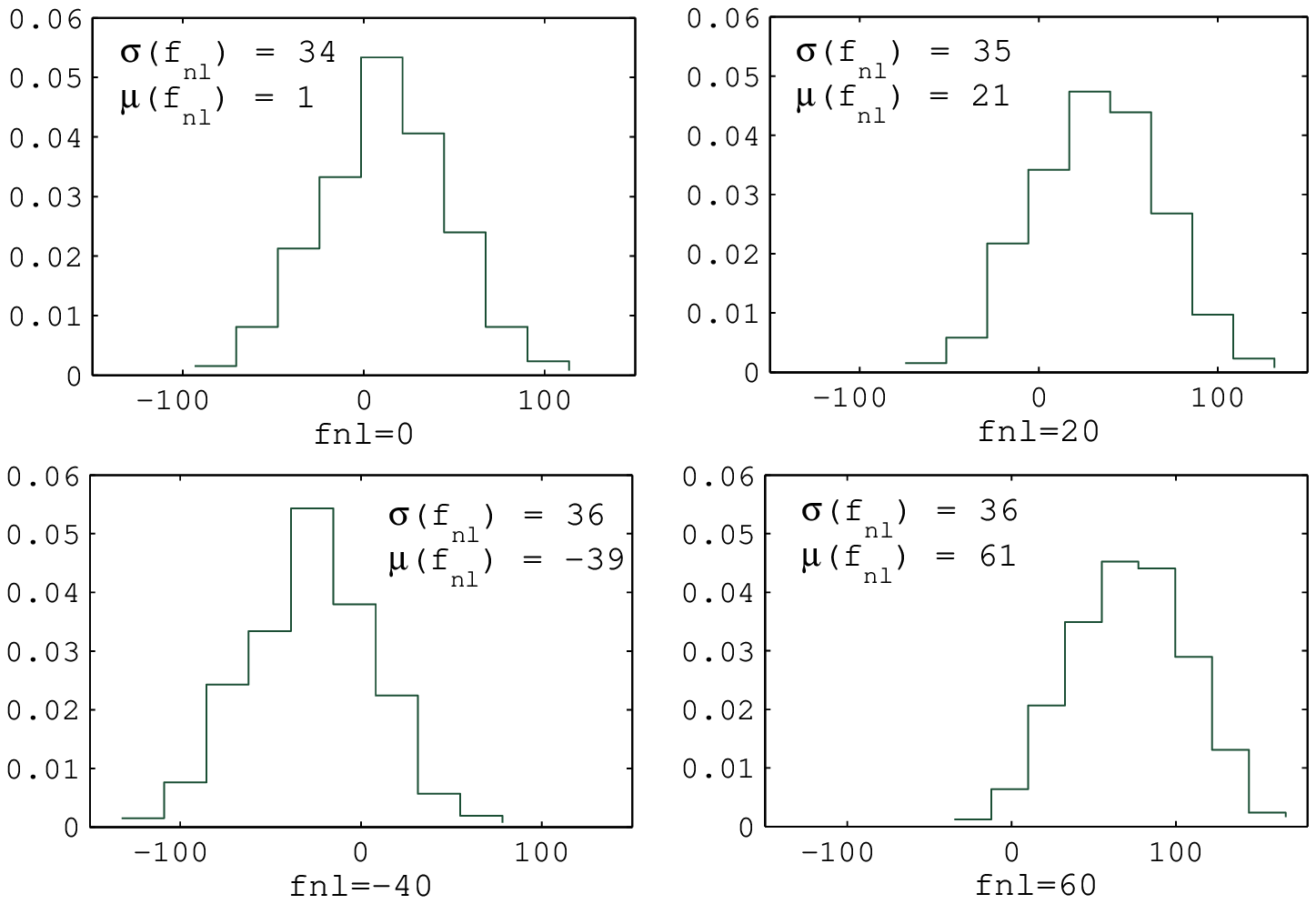} &\includegraphics[width=9cm,height=6cm]{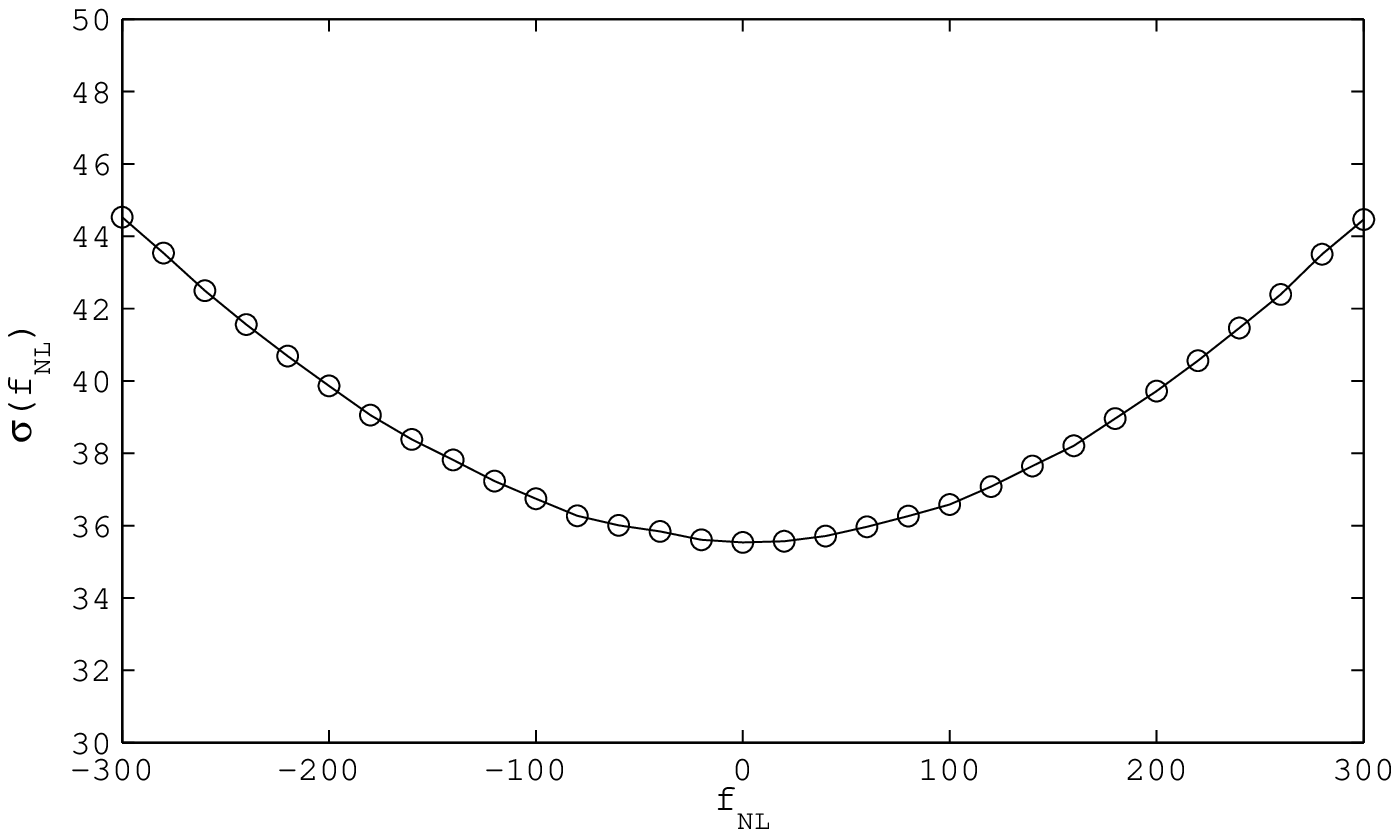} 
\end{tabular}
 \caption{The left part of the figure shows the histograms of the estimated $f_{NL}$ from simulations with values of $f_{NL}=-40,0,20,60$. The mean value and dispersion of $f_{NL}$ for each considered case is indicated in the corresponding panel. The right panel shows the behaviour of $\sigma(f_{NL})$ when estimated from simulations with different values of $f_{NL}$.} \label{fig:histo_fnl}
 \end{figure*}

In order to study further the robustness of our estimator, we have performed some additional tests. In particular, we have estimated the mean value and dispersion of the best-fit value of $f_{NL}$ from simulations with different values of $f_{NL}$. The left panel of Fig.~\ref{fig:histo_fnl} shows the histograms of the estimated values of $f_{NL}$ obtained from simulations with $f_{NL}$=-40,0,20,60. To carry out these tests, we have used 500 of the 1000 non-Gaussian simulations to estimate the mean value of the third order statistics $\langle v_{i}\rangle_{f_{NL}}$ and the remaining 500 simulations to obtain estimates of $f_{NL}$ and construct the histograms. The mean values and dispersions of $f_{NL}$ are given in the corresponding panels. In particular, we see that the method is unbiased, since the mean value of the estimated $f_{NL}$ is very close to the true underlying value for all the considered cases. In addition, we also plot in the right panel of Fig.~\ref{fig:histo_fnl} how the dispersion of the estimator varies as a function of $f_{NL}$. The standard procedure to estimate this dispersion is to use Gaussian simulations  but, as seen in the plot, this gives a minimum in the estimated value of $\sigma(f_{NL})$. However, for small values of $f_{NL}$, such as the ones found in this paper, the variation is small and therefore one can safely use the value of the dispersion estimated for the Gaussian case. 
 
Finally, we have repeated the same analysis using a set of 300 non-Gaussian simulations generated by \citet{Liguori2007}, finding very similar constraints on $f_{NL}$. 

\subsection{Point source contribution}
\begin{figure}
 \center
\includegraphics[width=9cm,height=6cm, angle = 0] {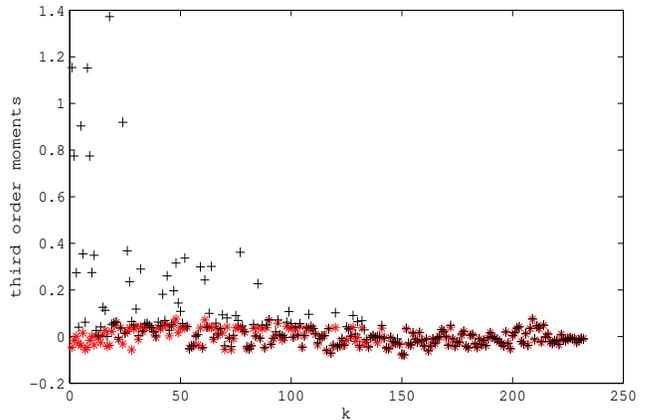} 
\caption{Mean values of the $v_{i}$ statistics obtained from 1000 Gaussian simulations with and without point sources. Diamonds represent CMB plus noise simulations, while crosses correspond to simulations including also the point sources. To improve the visualization, the statistics have been normalised to unit dispersion.}
 \label{fig:point_sources}
 \end{figure}

The background of unresolved point sources may introduce a bias in the estimation of $f_{NL}$. In order to correct this bias, we have studied the contribution to $f_{NL}$ given by a point source background that is added to the Gaussian simulations. For that purpose, we have produced point source simulations following the procedure of \citet{Curto2009a}. In particular, point sources maps are simulated according to the density distribution given by \citet{deZotti2005} in a range of intensities between $I_{min}=1$mJy and $I_{max}=1$Jy. These maps are then convolved with the corresponding beam and pixel functions and added to the simulations containing Gaussian CMB plus noise. The estimated value of $f_{NL}$ when point sources are present is then compared to the one obtained from simulations containing only CMB and noise, finding a difference of $\Delta f_{NL}=7\pm 6$. Fig.~\ref{fig:point_sources} shows the effect that point sources have on the $v_i$ statistics. As one would expect, they mainly affect the statistics involving small scales, that correspond to the ones with a lower value of $k$ in the figure.
Taking into account this result our final constraint on $f_{NL}$ for the WMAP-7yr data is $-69<f_{NL}<65$ at the 95 per cent confidence level.

\subsection{Local study of $f_{NL}$}
Finally, we have analysed the data considering two independent hemispheres. In particular, we have considered the hemispheres associated to the dipolar modulation proposed by \citet{Hoftuft2009} where the preferred direction is pointing towards the Galactic coordinates (l,b)=($224^{\circ}$,$-22^{\circ}$). We have estimated the best-fit value and constraints on $f_{NL}$ for the WMAP 7-yr data in both hemispheres, following the same procedure as for the full-sky. After correcting the point source contribution, the constraint found for the northern hemisphere is $-73<f_{NL}<119$ while for the southern hemisphere we have $-137<f_{NL}<62$, both at the 95 per cent confidence level. Therefore, as it was the case for the full-sky, both hemispheres are consistent with Gaussianity (i.e, $f_{NL}=0$). We have also tested that the results from the hemispheres are consistent between them. In particular, we have obtained the mean difference and dispersion between the $f_{NL}$ estimates at each hemisphere for Gaussian simulations, finding values of $\langle \Delta f_{NL} \rangle$=-4 and $\sigma (\Delta f_{NL})$=71. For the WMAP data, we have $\Delta f_{NL}=67$, which is perfectly consistent with the values expected from Gaussian simulations. Therefore, we do not find any assymmetry for the considered hemispheres. These results are in agreement with the analysis based on needlets made by \citet{Pietrobon2010} and \citet{Rudjord2010} for the WMAP-5yr data, where several divisions of the CMB map are studied without finding a significant asymmetry. In a recent work, \citet{Vielva2010} have found an asymmetry in the same hemispheres studied in this paper on the estimation of the $f_{NL}$ using the N-pdf. The disagreement may be caused by the differences on the methods. While \citet{Pietrobon2010} and \citet{Rudjord2010} have worked with the same resolution as we did (6.9 arcmin), \citet{Vielva2010} focused on scales around $2^{\circ}$. Also, the non-Gaussian model used by the former works is the same as the one used in this paper, whereas the model of the latter stands on the Sachs-Wolfe regime.     
%
%
\section{Conclusions}
\label{sec:conclusions}
We have presented a new methodology to analyse the Gaussianity of the CMB and to constrain the $f_{NL}$ parameter using the so-called HEALPix wavelet. To our knowledge, the developed $f_{NL}$ estimator is the fastest method that has been proposed up to date. In particular, for WMAP resolution ($N_\mathrm{side}=512$), it is $\sim 10^2$ times faster than the SMHW, $\sim 10^3$ times faster than the KSW bispectrum and $10^7$ times faster than the general bispectrum estimator. Moreover, although the constraints imposed by our method are not as stringent as those of the optimal estimators (based on the bispectrum or on the SMHW), they are very similar or even better than those proposed by alternative methods, such as needlets, Minkowski functionals or the N-pdf.

The method, which is based on the calculation of the third-order moments of the wavelet coefficient maps, has been applied to the WMAP-7yr V+W combined map. On the one hand, we have performed a $\chi^{2}$ test to study the Gaussianity of the CMB, finding consistency with the Gaussian hypothesis. On the other hand, we have constrained the value of the local $f_{NL}$ parameter to be $-69<f_{NL}<65$ at the 95 per cent confidence level, after correcting for the point source contribution. In addition, the HEALPix wavelet gives the possibility of performing local studies of Gaussianity in the CMB map. In particular, we have analysed two independent hemispheres associated to the dipolar modulation proposed by \citet{Hoftuft2009}. In this study, we do not find any significant asymmetry on the $f_{NL}$ estimates for the two hemispheres of the WMAP data. The constraints for the northern and southern hemispheres are $-73<f_{NL}<119$ and $-137<f_{NL}<62$, respectively, at the 95 per cent confidence level.     

\section*{acknowledgments}
We thank R.~Fern\'andez-Cobos for useful discussion on the HEALPix wavelet properties. We also thank M.~Liguori and F.~Elsner for help with the non-Gaussian simulations. The authors thank L.~Cabellos for computational support. We acknowledge partial financial support from the Spanish Ministerio de Ciencia e Innovaci\'on project AYA2007-68059-C03-02. B.~Casaponsa thanks the Spanish
Ministerio de Ciencia e Innovaci\'on for a pre-doctoral fellowship. P.~Vielva acknowledges financial
support from the Ram\'{o}n y Cajal programme. The authors acknowledge the computer resources, technical
expertise and assistance provided by the Spanish Supercomputing Network (RES) node at Universidad de Cantabria. We acknowledge the use
of Legacy Archive for Microwave Background Data Analysis (LAMBDA). Support for it is provided by the NASA Office of Space
Science. The HEALPix package was used throughout the data analysis \citep{Gorski2005}. 
\bibliographystyle{mn2e}
\bibliography{fnl_constraints_casaponsa_1s_revision}

\end{document}